\documentclass[12pt]{article}
\usepackage{graphicx}

\def\Re{{\cal R \mskip-4mu \lower.1ex \hbox{\it e}\,}}
\def\Im{{\cal I \mskip-5mu \lower.1ex \hbox{\it m}\,}}
\def\ie{{\it i.e.}}
\def\eg{{\it e.g.}}
\def\etc{{\it etc}}

\def\sub#1{_{\lower.25ex\hbox{$\scriptstyle#1$}}}
\def\tev{\,{\ifmmode\mathrm {TeV}\else TeV\fi}}
\def\gev{\,{\ifmmode\mathrm {GeV}\else GeV\fi}}
\def\mev{\,{\ifmmode\mathrm {MeV}\else MeV\fi}}
\def\mpl{\ifmmode M_{pl}\else $M_{pl}$\fi}
\def\mpl{\ifmmode \overline M_{Pl}\else $\bar M_{Pl}$\fi}
\def\to{\rightarrow}

\def\subw{_{\rm w}}
\def\mh{\ifmmode m\sbl H \else $m\sbl H$\fi}
\def\mch{\ifmmode m_{H^\pm} \else $m_{H^\pm}$\fi}
\def\mt{\ifmmode m_t\else $m_t$\fi}
\def\mc{\ifmmode m_c\else $m_c$\fi}
\def\mz{\ifmmode M_Z\else $M_Z$\fi}
\def\mw{\ifmmode M_W\else $M_W$\fi}
\def\mws{\ifmmode M_W^2 \else $M_W^2$\fi}
\def\mhs{\ifmmode m_H^2 \else $m_H^2$\fi}   
\def\mzs{\ifmmode M_Z^2 \else $M_Z^2$\fi}
\def\mts{\ifmmode m_t^2 \else $m_t^2$\fi}
\def\mcs{\ifmmode m_c^2 \else $m_c^2$\fi}
\def\mchs{\ifmmode m_{H^\pm}^2 \else $m_{H^\pm}^2$\fi}
\def\ztwo{\ifmmode Z_2\else $Z_2$\fi}
\def\zone{\ifmmode Z_1\else $Z_1$\fi}
\def\mtwo{\ifmmode M_2\else $M_2$\fi}
\def\mone{\ifmmode M_1\else $M_1$\fi}
\def\tb{\ifmmode \tan\beta \else $\tan\beta$\fi}
\def\xw{\ifmmode x\subw\else $x\subw$\fi}
\def\ch{\ifmmode H^\pm \else $H^\pm$\fi}
\def\lum{\ifmmode {\cal L}\else ${\cal L}$\fi}
\def\inpb{\,{\ifmmode {\mathrm {pb}}^{-1}\else ${\mathrm {pb}}^{-1}$\fi}}
\def\infb{\,{\ifmmode {\mathrm {fb}}^{-1}\else ${\mathrm {fb}}^{-1}$\fi}}
\def\epem{\ifmmode e^+e^-\else $e^+e^-$\fi}
\def\ppb{\ifmmode \bar pp\else $\bar pp$\fi}
\def\bsg{\ifmmode B\to X_s\gamma\else $B\to X_s\gamma$\fi}
\def\bsll{\ifmmode B\to X_s\ell^+\ell^-\else $B\to X_s\ell^+\ell^-$\fi}
\def\bstt{\ifmmode B\to X_s\tau^+\tau^-\else $B\to X_s\tau^+\tau^-$\fi}
\def\lamt{\ifmmode \tilde\lambda\else $\tilde\lambda$\fi}
\def\shat{\ifmmode \hat s\else $\hat s$\fi}
\def\that{\ifmmode \hat t\else $\hat t$\fi}
\def\uhat{\ifmmode \hat u\else $\hat u$\fi}

\newskip\zatskip \zatskip=0pt plus0pt minus0pt
\def\matth{\mathsurround=0pt}
\def\lsim{\mathrel{\mathpalette\atversim<}}

\def\atversim#1#2{\lower0.7ex\vbox{\baselineskip\zatskip\lineskip\zatskip
  \lineskiplimit 0pt\ialign{$\matth#1\hfil##\hfil$\crcr#2\crcr\sim\crcr}}}

\def\grtsim{\,\,\rlap{\raise 3pt\hbox{$>$}}{\lower 3pt\hbox{$\sim$}}\,\,}
\def\lsim{\,\,\rlap{\raise 3pt\hbox{$<$}}{\lower 3pt\hbox{$\sim$}}\,\,}

\renewcommand{\thefootnote}{\fnsymbol{footnote}}

\begin{document} \begin{titlepage}
\rightline{\vbox{\halign{&#\hfil\cr
&SLAC-PUB-11150\cr
&June 2005\cr}}}
\begin{center}
\thispagestyle{empty} \flushbottom { {\Large\bf Lorentz 
Violation in Extra Dimensions 
\footnote{Work supported in part
by the Department of Energy, Contract DE-AC02-76SF00515}
\footnote{e-mail:
rizzo@slac.stanford.edu}}}
\medskip
\end{center}

\centerline{Thomas G. Rizzo}
\vspace{8pt} 
\centerline{\it Stanford Linear
Accelerator Center, 2575 Sand Hill Rd., Menlo Park, CA, 94025}

\vspace*{0.3cm}

\begin{abstract}
In theories with extra dimensions it is well known that the Lorentz invariance 
of the $D=4+n$-dimensional spacetime is lost due to the compactified 
nature of the $n$ dimensions leaving invariance only in 4d. In such 
theories other sources of Lorentz violation may exist associated with the 
physics that initiated the compactification process at high scales. Here we 
consider the possibility of capturing some of this physics by analyzing the 
higher dimensional analog of the model of Colladay and Kostelecky. In that 
scenario a complete set of Lorentz violating operators arising from 
spontaneous Lorentz violation, that are not obviously Planck-scale suppressed,  
are added to the Standard Model action. Here we consider the influence of the 
analogous set of operators which break Lorentz invariance in 5d within the 
Universal Extra Dimensions picture. We show that such operators can greatly 
alter the anticipated Kaluza-Klein(KK) spectra, induce electroweak symmetry 
breaking at a scale related to the inverse compactification radius, yield 
sources of parity violation in, \eg, 4d QED/QCD and result in significant 
violations of KK-parity conservation produced by fermion Yukawa couplings, 
thus destabilizing the lightest KK particle. LV in 6d is briefly discussed. 
\end{abstract}



\renewcommand{\thefootnote}{\arabic{footnote}} \end{titlepage} 

%
%
%
%
%

\section{Introduction}

The possibility that Lorentz invariance(LI) may be violated at high 
energies in 4d with testable consequences has become a subject of much 
interest in the past few years{\cite {Matt}}. If one considers 
rotational invariance to be sacrosanct due to the strength of existing 
experimental constraints, Lorentz violation(LV) must take the form of a 
breakdown of invariance under boosts. Such a scenario is suggestive of 
spontaneous LV{\cite {spon}} where in some preferred frame, \eg,  a time-like 
4-vector takes on a vacuum expectation value with components 
$\sim (1,0,0,0)$. 

While LV of the type described above have yet to be observed in 4d it is clear 
that if $n$ extra (flat) compact dimensions exist, perhaps near the TeV 
scale, they obviously behave  
differently than do the large dimensions with which we are familiar. Clearly 
the LI in the $D=4+n$ dimensional space  has been lost 
through the compactification process; LI suffers further 
damage if the compactified manifold (here considered to be a torus) is 
orbifolded. In the usual bottom-up analysis, the $D$-dimensional 
action is conventionally 
written in a completely LI manner with all LV arising from the 
compactification/orbifolding process in the IR. Thus, in the UV, such a theory 
apparently 
maintains LI. While such an approach may be the simplest one to analyze it 
certainly does not address the larger issue as to how or why the $n$ 
dimensions have become compact. Some unknown UV physics must have 
triggered this compactification process making the 4 large dimensions (or, 
depending on one's viewpoint, the $n$ compactified 
dimensions) `special' otherwise we would be living in $4+n$ dimensions. 
Thus the true UV physics cannot be completely LI in $D$-dimensions.  
It would be interesting to ask if this UV physics has 
left any signatures for us to find at accelerators that are beginning to probe 
the TeV energy regime near the compactification 
scale, $R^{-1}$. If present, how might such effects be parameterized?  

For simplicity let us consider the case of one extra dimension, \ie, 
a theory in 5d with the 
extra dimension compactified as usual on $S^1/Z_2$ with $R^{-1} \lsim$  few 
TeV. If the UV breaking of 
LI is spontaneous, as in the 4d case discussed above, we can imagine that 
it was triggered by a 5-vector taking on the vev  $\sim (0,0,0,0,1)$ in 
some frame{\cite {spon}} 
thus leaving us with 4d LI with the fifth dimension 
becoming `special'. Although our approach will not capture any of the 
detailed dynamics associated with how such a vev was generated, 
it may be able to probe some of the residual LV physics which could  
remain accessible at the TeV scale. 
A framework for this type of analysis already exists for the Standard 
Model(SM) in 4d, the so-called SM Extension{\cite {Kost}}. This framework 
is particularly appealing for a number of reasons one of which is that it 
isolates the dominant effects of explicit LV in operators of the lowest possible 
dimension. In order to have a specific model in which to work in 5d 
that is closely analogous to this 4d case, we here adapt 
this particular framework to the 5d version of SM where all fields are in 
the bulk, \ie, the Universal Extra Dimensions(UED){\cite {UED}} scenario, but 
allowing LV to occur only in the fifth dimension. 

As we will see in detail below the existence of LV in 5d induces a number of 
new effects within the 5d UED framework such as: ($i$) The KK spectra of 
the gauge and Higgs bosons as well as those for all of the left- 
and right-handed fermions can each 
be rescaled arbitrarily. This can increase the possible confusion between 
UED and SUSY scenarios at the LHC{\cite {UED}}. ($ii$) New loop-induced parity 
violating effects are generated within previously parity conserving sectors 
of the model, \eg, in 5d QED/QCD. One signature of this is the generation of 
anapole moment-type  
couplings between fermions and gauge bosons. ($iii$) KK-parity, 
which is exact even at loop order in ordinary UED, becomes broken through 
mixings in the fermion KK mass matrices. This results in loop-induced mixing 
among gauge KK states and a finite lifetime for the Lightest KK Particle(LKKP) 
which usually considered as a stable dark matter candidate in the UED 
scenario. ($iv$) It is possible that the 5d LV operators may be the source 
of electroweak symmetry breaking. 

The outline of this paper is as follows: In Section 2 we discuss the 
adaptation of the LV SM Extension operators to 5d under the assumption that 
LI is broken only in 5d. Here we also show how the 5d analogs of the 4d 
CPT violating (as well as LV) operators can be (almost) removed from the action 
by suitable field redefinitions. We will show that through these redefinitions 
these operators may induce 
spontaneous symmetry breaking by generating a negative mass squared for 
scalar fields. This leads to a potential correlation between the SM Higgs 
vev and the size of the extra dimension, $R^{-1}$. 
In Section 3, we analyze how the remaining 
operators lead to modifications in the KK spectra of the SM gauge, scalar 
and fermion fields. We show that having a different KK spectrum for left-
and right-handed fermions, which is possible now that 5d LI is broken, yields 
a potential source for parity violation in QED/QCD in 4d. In Section 4 we 
demonstrate that KK-parity violation results from the Yukawa sector of the 
theory that normally generates masses for the would be fermion zero-modes. 
This again arises from the field redefinitions employed earlier to remove the 
analogs of the CPT violation operators. KK-parity violation is shown to lead 
to a number of new effects such as the instability of the lightest KK-parity 
odd particle as well as general mixing among the KK-even and KK-odd 
excitations. A discussion and our conclusions can be found in Section 5. 
The Appendix contains a brief discussion of LV in 6d for scalar fields.

\section{Lorentz Violating Operators}

Given the field content of the SM, the SM Extension{\cite {Kost}} provides 
for us a relatively 
short list of the lowest-dimension gauge invariant 4d 
LV operators which may also be CPT violating. We can easily adapt this list to 
our purposes and restrict ourselves to those cases where only 5d LI is lost 
while 4d LI remains. 
This requirement turns out to be highly restrictive as many of the 4d LV 
operators do not have 5d analogs which can lead to loss of LI in only 5d. 
In principle, one can extend this list by including new operators generated by 
gravity modifications in 5-d as has been done in Ref.{\cite {gravi}}. However, 
since we are restricting ourselves to the UED framework with one additional 
TeV scale extra dimension wherein gravitational effects we will not discuss this 
possibility here.

Systematically going through the list 4d operators we find a number whose 
generalizations to 5d cannot satisfy our constraints. 
Consider for example the LV set of 4d operators 
\begin{equation}
\Phi^\dagger(\alpha_{\mu\nu} B^{\mu\nu}+\beta_{\mu\nu}T^aW_a^{\mu\nu})\Phi
\,,
\end{equation}
where $\Phi$ is the Higgs scalar and $B^{\mu\nu}$ and $W_a^{\mu\nu}$ are 
the $U(1)$ and $SU(2)$ field strength tensors; $\alpha_{\mu\nu}$ and 
$\beta_{\mu\nu}$ are sets  
of numerical coefficients. Generalizing to 5d we immediately obtain
\begin{equation}
\Phi^\dagger(\alpha_{MN} B^{MN}+\beta_{MN}T^aW_a^{MN})\Phi\,. 
\end{equation}
We now ask what values of $A,B$ are allowed for the coefficients above 
without violating 4d LI: if $A,B$ both 
take on 4d indices then 4d LI will be broken. Similarly, if, \eg, $A=\mu$ and 
$B=5$ then 4d LI is again lost; the last possibility, \ie, $A=B=5$, yields 
zero due to the index antisymmetry. Thus the generalization of operators 
such as this in 4d to 5d does not yield anything interesting given the 
assumptions of our analysis. As another example of this, 
consider the 4d operator 
\begin{equation}
\kappa_{\mu\nu}\bar D \sigma^{\mu\nu}S\Phi+h.c.\,,
\end{equation}
where $D(S)$ is an $SU(2)_L$ doublet(singlet) fermion field and 
$\kappa_{\mu\nu}$ a set of numerical coefficients. In 5d this trivially 
generalizes to 
\begin{equation}
\kappa_{AB}\bar D \sigma^{AB}S\Phi+h.c.\,. 
\end{equation}
As in the previous example the various possible 
choices of $A,B$ yield either LV in 
4d or are zero by the antisymmetry of the indices. 

Let us now turn to the set of surviving operators. 
As an example, in 4d, suppressing flavor indices one has the following 
possible `kinetic' LV term, \eg, for an $SU(2)_L$ singlet fermion field, $S$:
\begin{equation}
{i\over {2}} (c_S)_{\mu\nu} \bar S\gamma^\mu \bar D^\nu S\,,
\end{equation}
where $\bar D^\nu$ is a covariant derivative acting in both directions and 
the $c_S$ are a set of dimensionless numerical 
coefficients; we expect these coefficients to be very small in 4d. Here we 
wish to generalize this operator to 5d, \ie, 
\begin{equation}
{i\over {2}} (c_S)_{AB} \bar S\Gamma^A \bar D^B S\,,
\end{equation}
where $\Gamma^\mu=\gamma^\mu$, $\Gamma^5=i\gamma_5$ 
and only keep terms which are LV in 5d but not in 4d. Clearly, given the 
experience of the examples above, we can only 
choose $A=B=5$ and taking $k_S=(c_S)_{55}$ this term becomes 
\begin{equation}
{k_S\over {2}}[\bar S\gamma_5 D_5 S-(D_5 \bar S) \gamma_5 S]\,. 
\end{equation}
Perhaps, more naturally, in 5d we might anticipate that $k_S =O(1)$ since LI 
in 4d remains unbroken. Of 
course we may expect a similar term to be present for an $SU(2)_L$ doublet 
as in the singlet case described 
above, with $k_S \to k_D$, but which need not be the same value. We will 
assume for simplicity that these 5d fermion 
terms are flavor-diagonal in what follows 
and denote their set of coefficients more generically by $k_\Psi$. 
Going through and attempting to generalize the remaining 
set of SM Extension 4d operators we find that only very few of them  
satisfy our 5d requirements; in addition to the `kinetic'-like operator above 
for fermions we find the following possibilities:
\begin{eqnarray}
{1\over {4}}k_{\kappa \lambda \mu \nu}F^{\kappa\lambda}F^{\mu\nu}&\to& 
{\lambda \over {4}}\Big(F_{\mu 5}F^{\mu 5}+F_{5\mu}F^{5\mu}\Big)\nonumber \\
k'_{\mu\nu} (D^\mu \Phi)^\dagger (D^\nu \Phi) &\to& -k_\Phi (D_5\Phi)^\dagger
(D_5\Phi)\nonumber \\
a_\mu \bar f\gamma^\mu f &\to& i\alpha \bar f \gamma_5 f\nonumber \\
i(k_\phi)^\mu \Phi^\dagger D_\mu \Phi +h.c. &\to& ih_\Phi \Phi^\dagger D_5 \Phi
+h.c.\,,
\end{eqnarray}
where the `mapping' from 4d to 5d is shown explicitly. In the equation above, 
$\Phi$ represents the Higgs doublet as before 
and, correspondingly, $F$ represents any 
of the SM field strength tensors. Likewise, $f$ is either an $SU(2)_L$ 
doublet, $D$, or singlet, $S$, fermion. Note that the first two operators 
above lead to modifications of gauge and Higgs 
kinetic terms as was the case for the fermionic operator discussed previously. 
Also note that in 4d the last two operators 
in the equation above are CPT violating; we note that in 5d their 
coefficients must be $Z_2$-odd in a manner similar to that of any  
5d bulk fermion 
mass term. The parameters $\lambda, k_{\Phi,\Psi}$ are dimensionless 
quantities which we might expect to be of order unity; they must be highly 
suppressed quantities in the usual 4d SM Extension. On the otherhand the 
coefficients $\alpha$ and $h_\Phi$ have the dimensions of mass and might most 
naturally be of order $\sim R^{-1}$. 
While it is possible that higher dimensional operators may also be present in 
addition to the ones considered above these are likely to be Planck suppressed 
and can be safely ignored in the analysis below. 

The LV operators that we have found above are for a 5d scenario. It would 
be interesting 
to consider how this operator set would be modified by going to even higher 
dimensions, \eg, 6d. Here we could imagine that not only is LI of the type 
that we have been discussing violated in the higher dimensional action but 
also rotational invariance in the extra dimensions may be lost leading 
to very interesting new physics. Such possibilities will be considered 
briefly in the Appendix and in detail elsewhere.  

It was noted in Ref{\cite {Kost}} that some of these CPT violating operators   
can be eliminated from the action 
in 4d by suitable field redefinitions. This remains 
especially true here in 5d (since we are only considering LV in the one extra 
dimension) but with interesting consequences since these field 
redefinitions will now depend on the co-ordinate of the extra fifth dimension. 
In a way, this field redefinition resums the effects of these operators 
in a non-perturbative way 
into the fields themselvess. Leaving these operators in the action, one would 
obtain similar effects order by order in the perturbation theory in the new LV 
parameters. The field redefinition simplifies the action and allows us to provide 
a non-perturbative treatment of these operators. 
As an example, consider the scalar part of the action including the two 
relevant LV terms above: 
\begin{equation}
\int d^4x ~dy ~\Big[(D_A \Phi)^\dagger (D^A \Phi)-V(\Phi^\dagger \Phi)-
k_\Phi(D_5 \Phi)^\dagger (D_5 \Phi)+ih_\Phi(\Phi^\dagger D_5 \Phi-\Phi D_5
\Phi^\dagger)\Big]\,,
\end{equation}
where we use $y$ as the co-ordinate for the extra dimension and $V$ is the 
usual scalar potential. If we now make a field redefinition 
\begin{equation}
\Phi \to e^{i\Sigma_\Phi y} \Phi\,,
\end{equation}
where $\Sigma$ is a parameter whose value we choose to be (recall $h_\Phi$ is 
$Z_2$-odd)  
\begin{equation}
\Sigma_\Phi=-{h_\Phi\over {1+k_\Phi}}\,, 
\end{equation}
then the action becomes 
\begin{equation}
\int d^4x ~dy ~\Big[(D_A \Phi)^\dagger (D^A \Phi)-V(\Phi^\dagger \Phi)+
\Sigma_\Phi^2(1+k_\Phi)\Phi^\dagger \Phi-k_\Phi(D_5 \Phi)^\dagger 
(D_5 \Phi)\Big]\,, 
\end{equation}
thus eliminating the `CPT violating' term but now introducing 
a new contribution to the scalar potential. Note that although 
the parameter $\Sigma_\Phi$ must be $Z_2$-odd to maintain the original symmetry  
only its square now appears in the action. Though the `CPT' violating operator 
no longer appears its effects will remain important as we will see below. 

Interestingly it is possible that this new quadratic term could 
produce a negative mass squared for the scalar Higgs field thus generating 
spontaneous symmetry breaking in the potential. 
Since we imagine that most naturally $\Sigma_\Phi \sim 
R^{-1}$ in magnitude this would tell us that the weak scale is linked to 
the size of the compactification scale up to order one corrections. 
Note that our field redfinition is consistent with the original $Z_2$ 
symmetry. Also note that additional kinetic term proportional to $k_\phi$ 
can induce tachyonic states, \ie, instabilities and /or causality violations, 
unless the value of this parameter is restricted to be $\geq -1$. Such 
effects were observed in the 4d case when LV was present{\cite {stab}}.

A similar field redefinition trick can also be applied in the fermion sector 
to rid ourselves of the `CPT violating' term. Let $\Psi$ be any 5d fermion 
field; the relevant action is then 
\begin{equation}
\int d^4x ~dy ~\Big[{i\over {2}}\bar \Psi \Gamma^A \bar D_A \Psi-{1\over {2}}
k_\Psi \bar \Psi \gamma_5 \bar D_5 \Psi -i\alpha \bar \Psi \gamma_5 \Psi\Big]
\,,
\end{equation}
where we have neglected any bulk mass terms as is standard in UED. Now we 
make the field redefinition (which maintains the original $Z_2$ symmetry) 
\begin{equation}
\Psi \to e^{i\Sigma_\Psi y} \Psi\,,
\end{equation}
with 
\begin{equation}
\Sigma_\Psi={\alpha\over {1+k_\Psi}}\,, 
\end{equation}
and the `CPT violating' term is eliminated leaving the action 
\begin{equation}
\int d^4x ~dy ~\Big[{i\over {2}}\bar \Psi \Gamma^A \bar D_A \Psi-{1\over {2}}
k_\Psi \bar \Psi \gamma_5 \bar D_5 \Psi\Big]
\,,
\end{equation}
this time with no additional terms. As in the case above we note that the 
coefficient $\Sigma_\Psi$ must be $Z_2$-odd. 
Thus out of the five possible LV 
structures in 5d we can eliminate two of them by field redefinitions; as we 
will see these redefinitions will return to haunt us later on. 
Note that the remaining LV terms are all modifications to kinetic terms 
and are all dimension-5, \ie, they are of the 
same dimension as are the usual SM-like terms in the 5d action.

\section{Influence of LV Terms: KK Spectrum}

The three remaining LV terms have a common feature: they are modifications 
of the 5d kinetic terms for fermions, Higgs bosons and gauge fields. They are 
analogous to (but not the same as) the addition of brane kinetic terms in the 
action{\cite {bt}} and will produce similar effects as we will now see. We 
remind the reader that the LV contributions discussed below occur at the tree 
level while the somewhat analogous effects observed in the UED occur at loop 
order. 

For simplicity let us first examine the case of the free scalar(Higgs) 
field; the action is then 
\begin{equation}
\int d^4x ~dy ~\Big[(\partial_A \Phi)^\dagger (\partial^A \Phi)
-\mu^2 \Phi^\dagger \Phi-k_\Phi(\partial_5 \Phi)^\dagger (\partial_5 \Phi)
\Big]\,, 
\end{equation}
where we have allowed a standard 
bulk mass term only for this discussion. Performing 
the Kaluza-Klein(KK) decomposition as usual 
\begin{equation}
\Phi=\sum_n \phi_n(x) \chi_n(y)\,,
\end{equation}
and imposing the orbifold boundary conditions one obtains the usual 
eigenfunctions $\chi_n(y)\sim \cos q_ny$ for $Z_2$-even fields where  
$q_n^2=m_n^2-\mu^2$, with $m_n$ being the KK mass. In addition, these 
wavefunctions also have the standard 
normalization $\int ~dy ~\chi_n(y) \chi_m(y)=\delta_{nm}$. 
However, the KK spectrum is now somewhat different than usual:
\begin{equation}
m_n^2=\mu^2+{n^2\over {R^2}}(1+k_\Phi)\,,
\end{equation}
where $R$ is the compactification radius and $n$ is an integer. The effect of 
the LV term is to rescale the KK excitation 
spectrum by some arbitrary amount. (Recall 
the we expect the dimensionless quantity $k_\Phi$ to be as large as 
order unity.) This is quite similar to the loop-induced 
radiative Higgs mass shift found in 
the case of UED induced by brane kinetic terms{\cite {UED}}. 
In that model the size of 
the contribution was logarithmically dependent on the cutoff but was under 
control numerically; here the rescaling occurs at the tree-level and 
is completely arbitrary. In order to 
insure a tachyon-free spectrum, \ie, stability, 
it is clear that we must have $k_\Phi >-1$.

Next we turn to the gauge fields; when the corresponding gauge group is not 
spontaneously broken, \eg, for the case of gluons in $SU(3)_c$, the action is 
\begin{equation}
\int d^4x ~dy ~\Bigg[-{1\over {4}} F_{AB}F^{AB}-
{\lambda_c \over {4}}\Big(F_{\mu 5}F^{\mu 5}+F_{5\mu}F^{5\mu}\Big)
\Bigg]\,, 
\end{equation}
where color indices have been suppressed. The KK decomposition is most 
conveniently performed in the physical $g_5=0$ gauge: 
\begin{equation}
g_\mu=\sum_n g_\mu^{(n)}(x) f_n(y)\,,
\end{equation}
and produces the standard eigenfunctions $\sim \cos ny/R$ for $Z_2$-even 
fields normalized as usual. In a manner similar to the scalar case above, the 
KK masses are, however, now given by 
\begin{equation}
m_{g_n}^2={n^2\over {R^2}}(1+\lambda_c)\,,
\end{equation}
where $\lambda_c$ can be O(1). This spectrum shift is again 
similar to that induced by brane term radiative corrections in the UED model 
but here it can rescale the spectrum arbitrarily by a large amount. Since 
$\lambda$ and $k_\Phi$ are completely unrelated, this rescaling of the KK 
spectra can be performed independently for these fields.   

In the electroweak sector the KK decomposition is a bit more complex due 
to presence of symmetry breaking, the mixing among the neutral fields and the 
fact that the LV coefficients for the $SU(2)_L$ and $U(1)_Y$ gauge groups, 
$\lambda_{W,B}$, respectively, can be numerically different. The case of the 
$W$ KK tower is rather straightforward since the effect of symmetry breaking 
here is to generate a simple bulk mass term with the usual eigenfunctions; one 
obtains in standard notation 
\begin{equation}
m_{W_n}^2={1\over {4}} g^2v^2+{n^2\over {R^2}}(1+\lambda_W)\,,
\end{equation}
as we might have expected. Note that we can adjust the $W$ and gluon towers 
relative to each other in an arbitrary manner; in UED the ratio of these two,  
loop-induced shifts is completely fixed. For the neutral fields one obtains a 
level-by-level mixing between the KK excitations of the $B$ and $W_3$ fields 
as in the case of UED. The elements of the symmetric KK mixing matrix at the 
$n^{th}$ level are given by
\begin{eqnarray}
M_{W_3W_3}^2&=&{1\over {4}} g^2v^2+{n^2\over {R^2}}(1+\lambda_W)\nonumber \\
M_{W_3B}^2=M_{BW_3}^2&=&{1\over {4}} gg'v^2\nonumber \\
M_{BB}^2&=&{1\over {4}}g'^2v^2+{n^2\over {R^2}}(1+\lambda_B)\,,
\end{eqnarray}
which is somewhat similar to the case of UED. The corresponding 
level-dependent `Weinberg-angle' is then given by 
\begin{equation}
\tan 2\theta_n = {-2gg'v^2\over {4(\lambda_B-\lambda_W){n^2\over {R^2}}+
(g'^2-g^2)v^2}}\,.
\end{equation}
Since $\lambda_{B,W}$ are arbitrary, in principle O(1) parameters, this KK 
mass spectrum for these two neutral fields 
can be substantially different than obtained in UED. 

The case for fermions proceeds in the standard fashion from the above action. 
Let us ignore $SU(2)_L\times U(1)_Y$ symmetry breaking and zero-mode mass 
generation for the moment; 
we will return to this issue below. The arbitrary 5d fermion 
field $\Psi$ is decomposed into left- and right-handed pieces in the 
standard manner: 
$\Psi=P_L\Psi_L+P_R\Psi_R$ using the usual projection operators and then 
the KK decomposition is performed, \ie, 
$\Psi_{L,R}=\sum_n \psi_{L,R}^{(n)}f_{L,R}^{(n)}(y)$ and we arrive at the 
the coupled equations
\begin{eqnarray}
(1+k_\Psi)\partial_y f_L^{(n)}&=& m_n f_R^{(n)}\nonumber \\
-(1+k_\Psi)\partial_y f_R^{(n)}&=& m_n f_L^{(n)}\,, 
\end{eqnarray}
so that the fermion masses are given by 
\begin{equation}
m_{\Psi_n}^2=(1+k_\Psi)^2 {n^2\over {R^2}}\,.
\end{equation}
For doublet(singlet) fields we choose the left(right)-handed fermions to be 
$Z_2$-even to obtain the conventional SM structure. 
Note that the mass-squared of the fermion fields are quadratic in the LV 
correction whereas bosons experience a linear correction. In either case we 
again see that we can rescale the mass spectrum as we'd like since we can 
choose the LV coefficients arbitrarily. In particular, there is no reason, 
\eg, for the left- and right-handed SM fermions to have KK towers that are in 
any way degenerate which can lead to new physics signatures as will be 
discussed below. Although KK-parity is still maintained at this point, 
clearly if one can 
rescale all of the masses of the KK excitations of the SM fields by arbitrary 
amounts it is no longer clear which state will be the lightest one in the 
spectrum. The identity of the LKKP dark matter candidate now depends on the 
values of the LV coefficients. We note that since we can rescale the fermion 
and boson spectrum as we'd like the possibility of confusion between the UED 
and SUSY scenarios at the LHC is now significantly increased.  

\begin{figure}[htbp]
\centerline{
\includegraphics[width=8.5cm,angle=90]{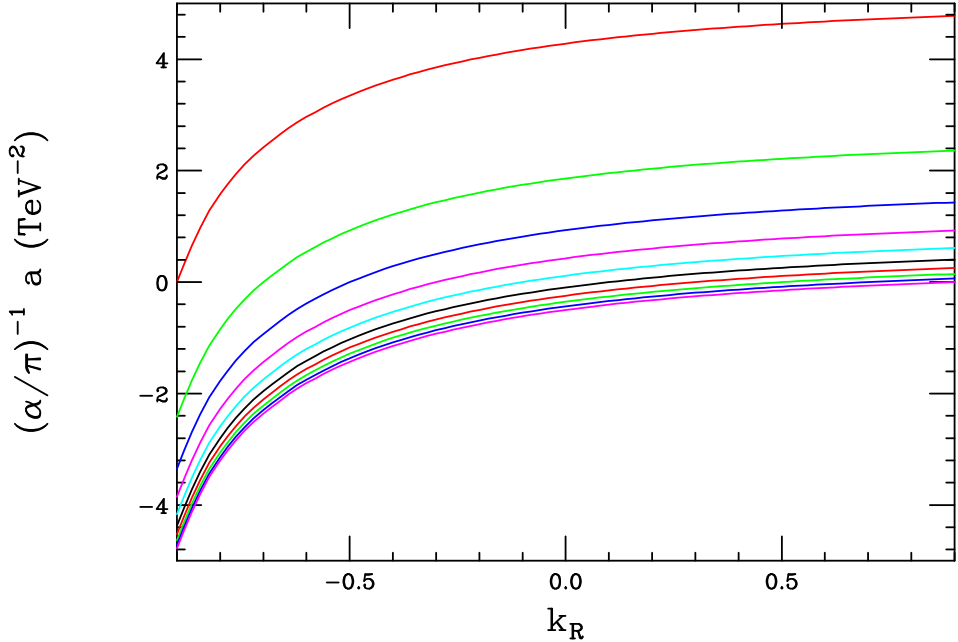}}
\vspace*{0.1cm}
\caption{The anapole moment of the electron induced by LV in 5d QED assuming 
$\lambda_\gamma=0$. From top to bottom the curves correspond to $k_L=$-0.9, 
-0.7,-0.5, \etc}
\label{fig1}
\end{figure}

The fact that the KK excitations of left-handed doublet and right-handed 
singlet fermions now have different tree-level 
masses directly leads to new phenomena. As a 
demonstration of this, consider for simplicity the toy model of 
5d QED accompanied by LV. The KK towers of the of the 
left-handed and right-handed electron now having different masses will  
produce 
a signal for {\it parity violation} within a conventionally parity conserving 
scenario as we will now demonstrate. If one considers the coupling between 
the (zero-mode) photon and left- and right-handed electrons one finds that 
at loop level a parity violating coupling will be generated, \ie, the 
anapole moment{\cite {ana}}, which corresponds to a tensor structure  
\begin{equation}
<f|J^{em}_\mu|f>_{anapole}=ieQ_f\bar f [q^2\gamma_\mu-\gamma\cdot qq_\mu]
\gamma_5 F_3(q^2)f\,,
\end{equation}
with $F_3(q^2)$ being the anapole moment form factor. $F_3(0)=a$ is then just 
the standard anapole moment which has dimensions $\sim R^{2}$. 
The existence of this 
coupling is directly related to the fact that the masses of the KK 
states inside the loop are different for the left- and right-handed towers; 
in obvious notation and summing over KK levels we 
find that
\begin{equation}
a\simeq {\alpha\over {\pi}}~{{\pi^2R^2}\over {48}}\int^1_0~dx \int^{1-x}_0~dy 
\Bigg[{{4(1-x)(1-y)+5xy}\over {1+\lambda_\gamma+[(1+k_R)^2-1-\lambda_\gamma]
(x+y)}}-(R\to L)\Bigg]\,. 
\end{equation}
Assuming that $R^{-1}=$500 GeV and $\lambda_\gamma=0$ for purposes of 
demonstration we obtain the numerical results for $a$ shown in 
Fig.~\ref{fig1}; here we see that $|a|\sim \alpha$ TeV$^{-2}$, which to set 
the scale, is comparable in magnitude to the conventional SM 
contribution{\cite {ana}} induced by the parity-violating weak interaction.
Clearly LV in 5d can lead to parity violating signatures in 4d in the absence 
of weak interactions. 
Within the 5d UED scenario the analysis above also leads directly to a QCD 
color-anapole moment which also violates parity in 4d. 

As we have just seen the introduction of 5d LV violating operators with 
O(1) coefficients allows us to modify the overall scales of the various 
gauge, scalar and left- and right-handed 
fermion KK spectra in UED in an independent fashion. Thus 
when such operators are present it is no longer clear that, \eg, a neutral 
field will be the lightest state which is odd under KK-parity and we may 
lose our natural dark matter candidate. The situation is actually more severe 
than this as we will now see.

\section{Influence of LV Terms: KK-Parity Violation}

The remaining term in the action that we have yet to examine is 
the Yukawa coupling 
between the fermion doublet and singlet fields and the Higgs boson that 
generates non-zero masses 
for the (would-be) zero-mode SM fermions. We can write this generically, 
dropping all generation labels, as: 
\begin{equation}
\int d^4x ~dy ~\lambda_5 \bar DS\Phi+h.c.\,. 
\end{equation}
After rescaling by the field redefinitions employed above to rid ourselves of 
the unwanted `CPT violating' LV terms this action becomes
\begin{equation}
\int d^4x ~dy ~\lambda_5 e^{i(\Sigma_\Phi+\Sigma_S-\Sigma_D)y}\bar 
DS\Phi+h.c.\,,  
\end{equation}
so that a position-dependent phase has crept into the Yukawa part of the 
action. It is important 
to recall that the quantities $\Sigma_i$ are $Z_2$-odd, \ie, they flip their 
sign at the origin. To probe the influence of this term let us first 
extract out the all zero-mode piece and perform 
the $y$-integration. Recall that zero-mode wave functions are 
flat $=1/\sqrt{2\pi R}$; we obtain the 4d integrand 
\begin{equation}
{\lambda_5\over {\sqrt {2\pi R}}} ~{{v+H}\over {\sqrt {2}}} 
~e^{i\sigma \pi R/2} ~{\sin(\sigma \pi R/2)\over{\sigma \pi R/2}}\,,  
\end{equation}
where $\sigma=\Sigma_\Phi+\Sigma_S-\Sigma_D$ and $H$ is the usual SM Higgs 
field. The SM 4d Yukawa coupling can then be identified as 
\begin{equation}
\lambda_4={\lambda_5\over {\sqrt {2\pi R}}}  
~e^{i\sigma \pi R/2} ~{\sin(\sigma \pi R/2)\over{\sigma \pi R/2}}\,.   
\end{equation}
Apart from the overall phase factor the last term can substantially rescale 
the size of the Yukawa coupling depending on the value of $\sigma R$ and may 
lead to some interesting phenomenology.

Something even more interesting results when we do not project into the 
all zero-mode state. Due to the additional $y$-dependent phase these Yukawa 
terms can {\it violate} KK-parity causing, \eg, a destabilization of the 
LKKP. Recall that in the usual UED model KK-parity is preserved to all orders 
in perturbation theory. 
To see this effect it is useful to examine the mixing between the 
would-be zero mode fermion and the $Z_2$-even members of the KK tower; this 
corresponds to the off-diagonal sub-matrix linking, \eg, the zero-mode 
doublet field with a KK singlet state. This 
calculation is straightforward and, in terms of the 4d Yukawa coupling 
$\lambda_4$ is given by 
\begin{equation}
{{\lambda_4 v}\over {2}} \Bigg[e^{in\pi/2}~{{\sigma R}\over {\sigma R+n}}~
{{\sin((\sigma R+n)\pi/2)}\over {\sin(\sigma \pi R/2)}}\Bigg] +(n\to -n)\,,   
\end{equation}
which corresponds to the $0n$ element of the KK mass sub-matrix, $M_{0n}$, 
and is seen to be proportional 
to the SM zero mode mass, $M_{0n}=\delta_{0n} m_f$. (It is important to note 
that here the symbol $\delta_{0n}$ does {\it not} denote the Kronecker delta.) 
Clearly, such terms can only be significant if $\sigma R$ is O(1) but this 
might be expected. Furthermore, one finds that 
all of the sub-matrix elements of this type, $M_{nm}=m_f\delta_{nm}$, are 
in general found to be non-zero with a mass scale set 
by the conventional SM fermion mass, \ie, with $\delta_{nm}$ values 
generally of order unity and controlled by the values of $n,m$ and $\sigma R$. 
This is unlike the case of UED where the mixing 
between the $D$ and $S$ fermion fields takes place level by level; here there 
is also a potentially significant mixing between the various KK levels. 
However,  light fermions, such as  the electron, experience 
little direct KK-parity violation through such mixing 
whereas for heavy fields, like the top 
quark, this violation can be quite significant for $R^{-1}\sim 1$ TeV or less.
The removal of the `CPT-violating' LV terms in the original action via field 
redefinitions has thus resulted in the breakdown of KK-parity conservation. 
We note that 
KK-parity violation at some level might also occur if UED is extended higher 
dimensions to 
include gravitational effects{\cite {new}}.

This violation of KK-parity is quantifiable at the tree level
by estimating the lifetime of 
the LKKP. To get an order of magnitude estimate we perform the calculation 
in the mass insertion approximation and assume that as usual 
the first KK photon 
excitation with mass $M$ is the LKKP. The process proceeds via 
$\gamma_1 \to \bar f_{L0} f_{L1}+h.c. \to \bar f_{L0} f_{L0}+(L\to R)$,  
where the 
second step arises from mixing. We obtain
\begin{equation}
\Gamma=N_cQ_f^2(2Re~\delta_{01})^2 ~{{\alpha M}\over {6}} 
~(1-4m_f^2/M^2)^{1/2}\Big[(g_L^2+g_R^2)\Big(1-{{m_f^2}\over {M^2}}\Big)+6g_Lg_R
{{m_f^2}\over {M^2}}\Big]\,,   
\end{equation}
where $m_f$ is the would-be zero mode mass, $M=M_{\gamma_1}$,  
\begin{equation}
g_{L,R}={m_f^2\over {(m_{L,R}^2-m_f^2)}}\,,
\end{equation}
with $m_{L(R)}$ 
being the mass of the first KK excitation of the doublet (singlet) field 
$f_{L(R)}$ and $\delta_{01}$ is the dimensionless mixing parameter defined 
above.

\begin{figure}[htbp]
\centerline{
\includegraphics[width=8.5cm,angle=90]{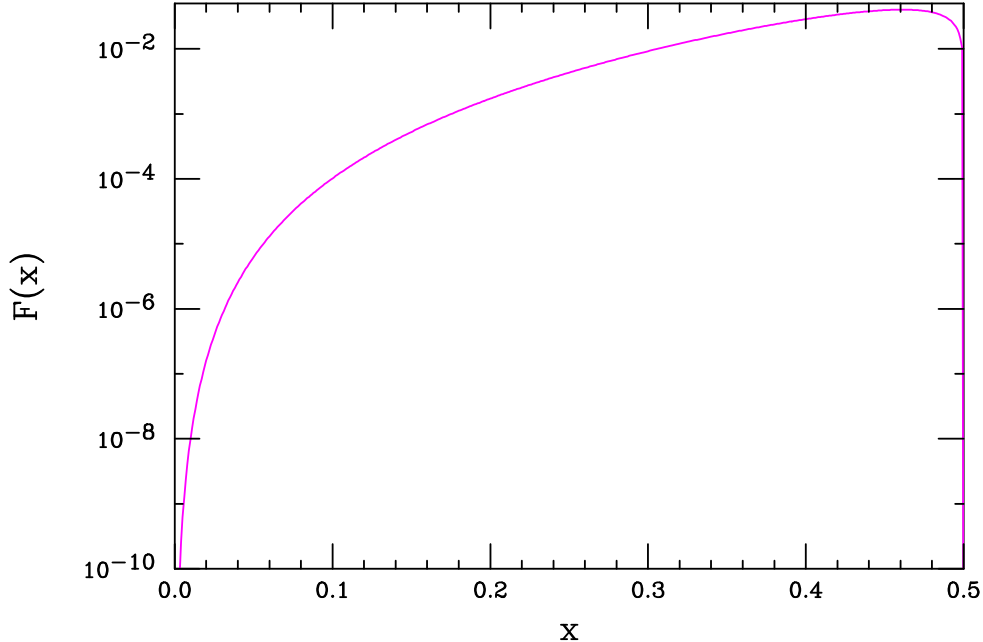}}
\vspace*{0.1cm}
\caption{The function F(x) as defined in the text.}
\label{fig2}
\end{figure}

Here we see again an example of induced parity violation in 
that the two couplings are equal, $g_L=g_R$, only when the 
fermion KK excitation masses are the same. Note that as expected this 
decay is very highly  
suppressed for light fermions, \ie, decays to heavy fermions such as  
top quarks, will be by far dominant.  
To get an idea of the size of this suppression, we take $m_L=m_R=M$ and 
$2Re(\delta_{01})=1$ so that
\begin{equation}
\Gamma={N_c\over {3}}Q_f^2\alpha ~M~F(x)\,,   
\end{equation}
with $x=m_f/M$;  the function $F(x)$ is shown in Fig.~\ref{fig2}. 
As we expected, except for 
the closure of phase space $F(x)$ is larger the closer $x$ is to 1/2; decays 
to first generation fields is thus seen to be highly suppressed. 
Although the expression above might correspond to a very narrow width 
by usual collider standards, 
for any reasonable range of parameters the lifetime of the LKKP is 
quite short in comparison to the age of the universe. Clearly, if some other 
particle is actually the LKKP, the analogous calculation can be performed 
obtaining qualitatively similar results. 

Another way to observe the violation of KK-parity is through 
loop-induced mixing among 
different gauge boson KK levels. This mixing is induced by the insertion of 
off-diagonal fermion mass matrix elements into vacuum polarization graphs 
connecting gauge fields with different KK number. In the 5d QED example 
this corresponds to a process $\gamma_n\to \bar f_n f_0+h.c.\to \bar f_m f_0 
+h.c.\to \gamma_m$ 
where the intermediate step occurs due to Yukawa induced fermion mixing. 
Using the notation above, mass mixing arising from this process in the photon 
tower mass matrix induced by a single fermion flavor is given by
\begin{equation}
\delta M_{mn}^2 \simeq N_c Q_f^2 ~{\alpha \over {\pi}}~2Re(\delta_{mn}) ~m_f^2
~G({m_f^2\over {m_{S_n}^2}},{m_f^2\over {m_{D_m}^2}})+(n\to m)\,,   
\end{equation}
with $m_{D_m,S_n}$ being the masses of the KK fermions in the loop and 
$G$  is an order one loop function. Here we again see 
that the dominant contribution arises from the most massive SM fermion sector 
as we might have expected. Clearly with $\delta_{nm}$'s of order unity 
a summation over all possible fermions in the loop can lead to small yet 
significant mixing in the gauge boson mass matrix. 
This result easily generalizes to the cases of the $W,Z$ and gluon KK towers 
where gauge KK mass eigenstates will now no longer have a definite KK-parity. 
A similar mixing will occur among Higgs and Goldstone KK levels.

One of the other 
effects of KK-parity conservation in the UED model is the inability 
to singly produce states which are KK-parity odd at colliders, \eg, the 
lightest KK 
gauge boson excitations. The violation of KK-parity induced by 
Yukawa interactions leads to modifications of this conventional result though 
the corresponding cross sections are not necessarily large. This can be seen 
by the fact that the widths of the KK-odd gauge bosons into zero modes 
of the first two generations is quite small.   

In this section we have seen that the elimination of the 5d analogs of the 4d  
`CPT violating' operators by field redefinitions induces potentially large 
violations of KK-parity. We observed that the size of this violation an any 
given SM fermion sector is correlated with the known size of the would-be 
zero mode masses. As a result UED loses its dark matter candidate.

\section{Summary and Conclusions}

In this paper we have initiated a study of the influence of explicit Lorentz 
violation within the context of the 5d SM where all fields are in the bulk, 
\ie, the Universal Extra Dimensions scenario. To perform this analysis we 
extended the `conventional' 4d model of Colladay and Kostelecky to 5d and 
searched for a subset of operators that can leave 4d Lorentz invariance 
untouched while breaking it in 5d. Two of these operators, the 5d analogs of 
those that violate CPT in 4d, can be (almost) removed from the action through 
a set of field redefinitions for fermions and scalars. One obvious result of 
this field redefinition is to induce a negative mass square term in the 
Higgs potential which may be the source of electroweak symmetry breaking. In 
addition, the natural scale of the induced vev would be $\sim R^{-1}$ thus 
linking the scale of electroweak symmetry breaking with the size of the extra 
dimension. The 
remaining LV operators lead to alterations of the various gauge, Higgs and 
fermionic kinetic terms and independently rescale their associated KK spectra 
which can increase the possible confusion of UED and SUSY at the LHC. 
Since, \eg, 
the masses of KK excitations of the left- and right-handed SM fermions need no 
longer be equal this induces, at loop order, parity-violating effects in 
previously parity-conserving parts of the SM, \ie, QED and QCD. Furthermore, 
we have shown that the the field redefinitions used to eliminate the 5d 
analogs of the 4d CPT violating terms make an important change in the 
nature of the Yukawa couplings responsible for generating the would-be 
zero-mode 
fermion masses. Due to an additional fifth co-ordinate-dependent phase, 
fermion mass terms are generated that produce mixing among all of the various 
KK levels thus violating KK-parity. This leads to a destabilization of the 
lightest KK-odd particle which is the usual dark matter candidate in UED. In 
addition these terms were shown to induce mixing between the various gauge 
KK levels at one-loop. 

As we have seen the presence of LV terms in the 5d UED scenario can lead to 
substantial modifications from the conventional expectations. Hopefully signals 
for extra dimensions will be found at future colliders.

\noindent{\Large\bf Acknowledgments}

The author would like to thank J.Hewett and B. Lillie for discussions 
related to this work.

\section*{Appendix: LV in 6d}

It is interesting to consider what happens when LV is extended to 6d and 
compactified on  
an orthogonal torus with radii $R_{5,6}$. Although a 
detailed study lies outside the scope of the current work we would like to give 
some flavor here by considering for simplicity the LV 6d scalar action. From 
the analysis above, this is given by 
\begin{equation}
\int d^4x ~dx_5~dx_6 ~\Big[(D_A \Phi)^\dagger (D^A \Phi)-V(\Phi^\dagger \Phi)-
k_{ij}(D_i \Phi)^\dagger (D_j \Phi)+ih_i(\Phi^\dagger D_i \Phi-\Phi D_i
\Phi^\dagger)\Big]\,,
\end{equation}
where we take $k_{i,j}$, $i,j=5,6$, to be real and symmetric; 
summation over these indices when repeated is implied. Since we 
will be concentrating for simplicity on the pure 
scalar sector in our discussion below, we have ignored the possibility of new 
LV interaction terms that may be present in 6d which are absent in 5d. 
As in 5d,  
the `CPT-violating' terms can be eliminated by a field redefinition: 
\begin{equation}
\Phi \to e^{i\Sigma_i x_i} \Phi\,,
\end{equation}
where
\begin{equation}
\Sigma_5={{-h_5(1+k_{66})+h_6k_{56}}\over {(1+k_{55})(1+k_{66})-k_{56}^2}}\,,
\end{equation}
and $\Sigma_6$ can be obtained by interchanging 5 and 6 in the expression 
above. As in 5d this field redefinition adds a new, likely negative term 
to the scalar potential:  
\begin{equation}
-{{h_5^2(1+k_{66})+h_6^2(1+k_{55})+2h_5h_6k_{56}}
\over {(1+k_{55})(1+k_{66})-k_{56}^2}}~\Phi^\dagger \Phi\,.  
\end{equation}
So far this is a rather straightforward extension of 5d; something new happens 
when we perform the usual KK decomposition
\begin{equation}
\Phi(x_\mu,x_i)=\sum_{n,m} \phi_{n,m}(x_\mu) \chi_{n,m}(x_i)\,. 
\end{equation}
Through the usual manipulations 
we are led to the equation of motion for $\chi$ which we can write for 
free scalars as 
\begin{equation}
\partial_i\Big[h^{ij}\partial_j \chi\Big]-m_{n,m}^2\chi=0\,,
\end{equation}
where the symmetric object $h_{ij}$ acts as a flat, constant 
`metric' in the $x_5-x_6$ space with elements  
$h_{55}=1+k_{55}$, $h_{66}=1+k_{66}$ and $h_{56}=k_{56}$. These satisfy 
$h_{il}h^{lj}=\delta_i^j$ and thus $h^{ij}$ are  
the elements of the inverse matrix $h^{-1}$. In the $x_5-x_6$ 
co-ordinate basis this equation is not generally 
separable; however, the metric can be 
diagonalized through a suitable $x_5-x_6$ rotation to the basis $x_\pm$: 
\begin{eqnarray}
x_+&=&x_5 \cos \theta+x_6 \sin \theta \nonumber \\
x_-&=&x_6 \cos \theta-x_5 \sin \theta \,,
\end{eqnarray}
with angle $\theta$ given by 
\begin{equation}
\tan 2\theta ={2k_{56}\over {k_{55}-k_{66}}}\,,
\end{equation}
so that the now separable equation of motion for $\chi$ becomes 
\begin{equation}
\lambda_+^{-1}\partial_+^2 \chi +\lambda_-^{-1}\partial_-^2 \chi 
+m_{n,m}^2\chi=0\,,
\end{equation}
with $\lambda_\pm$ given by 
\begin{equation}
\lambda_\pm=1+{{k_{55}+k_{66}}\over {2}} \pm {1\over {2}}\Big[(k_{55}-k_{66})^2
+4k_{56}^2\Big]^{1/2}\,. 
\end{equation}
Note that although our metric is constant, rotations no longer commute with it. 
The fact that there is a `preferred'  
frame where the `metric' is diagonal is the 
result of LV here manifest as the loss of $x_5-x_6$ rotational invariance.   
We can now express $\chi$ as $\chi_{n,m}=f_n(x_+)g_m(x_-)$ in this preferred 
basis. 

Although we have switched to the co-ordinates $x_\pm$, the boundary conditions 
will most likely be expressed in the $x_{5,6}$ basis. Here, for example, we 
consider the most simple case where we have invariance under the typical 
periodic conditions: 
$x_{5,6}\to x_{5,6}+2\pi R_{5,6}$, so that one can write 
$\chi =\exp in_ix_i/R_i= \exp i[a_+x_+ +a_-x_-]$ and thus  
\begin{equation}
m_{n_5,n_6}^2={a_+^2\over {\lambda_+}}+{a_-^2\over {\lambda_-}}\,,
\end{equation}
where
\begin{eqnarray}
a_+&=&{n_6\over {R_6}}\cos \theta -{n_5\over {R_5}}\sin \theta \nonumber \\
a_-&=&{n_5\over {R_5}}\cos \theta +{n_6\over {R_6}}\sin \theta \,. 
\end{eqnarray}
Note that in this simple case, the KK mass eigenvalue equation could have 
been obtained without making the co-ordinate transformation above since 
the eigenfunctions are simple exponentials. Straightforward algebra yields
\begin{equation}
m_{n_5,n_6}^2=\Bigg[(1+k_{55})(1+k_{66})+k_{56}^2\Bigg]^{-1} \Bigg((1+k_{66})
{{n_5^2}\over {R_5^2}}+(1+k_{55}){{n_6^2}\over {R_6^2}}-2k_{56}
{{n_5n_6}\over {R_5R_6}}\Bigg)\,.
\end{equation}
This eigenvalue equation for the KK masses is remarkably similar to that 
obtained by Dienes{\cite {Dienes}} who consider tori with shift angles and 
shape moduli in 6d. Instead 
of the simple KK spectrum rescaling that we observed for LV in 5d, in 6d the KK 
spectrum is significantly skewed and 
distorted compared to conventional expectations. The shift 
angle of Dienes in our case arises from LV and the existence of the  
preferred frame. 

A more detailed discussion of LV in 6d will be given elsewhere. 

%
\def\MPL #1 #2 #3 {Mod. Phys. Lett. {\bf#1},\ #2 (#3)}
\def\NPB #1 #2 #3 {Nucl. Phys. {\bf#1},\ #2 (#3)}
\def\PLB #1 #2 #3 {Phys. Lett. {\bf#1},\ #2 (#3)}
\def\PR #1 #2 #3 {Phys. Rep. {\bf#1},\ #2 (#3)}
\def\PRD #1 #2 #3 {Phys. Rev. {\bf#1},\ #2 (#3)}
\def\PRL #1 #2 #3 {Phys. Rev. Lett. {\bf#1},\ #2 (#3)}
\def\RMP #1 #2 #3 {Rev. Mod. Phys. {\bf#1},\ #2 (#3)}
\def\NIM #1 #2 #3 {Nuc. Inst. Meth. {\bf#1},\ #2 (#3)}
\def\ZPC #1 #2 #3 {Z. Phys. {\bf#1},\ #2 (#3)}
\def\EJPC #1 #2 #3 {E. Phys. J. {\bf#1},\ #2 (#3)}
\def\IJMP #1 #2 #3 {Int. J. Mod. Phys. {\bf#1},\ #2 (#3)}
\def\JHEP #1 #2 #3 {J. High En. Phys. {\bf#1},\ #2 (#3)}

\end{document}